\documentclass[prb,onecolumn,nofootinbib,citeautoscript,10pt,longbibliography,notitlepage]{revtex4-2}

\usepackage[utf8]{inputenc}
\usepackage{graphicx}
\usepackage{dcolumn}
\usepackage{bm}
\usepackage{color} 
\usepackage{amsmath}
\usepackage{tabularx,graphicx}
\usepackage{epstopdf}
\usepackage{latexsym}
\usepackage{amssymb}
\usepackage{color, colortbl}
\usepackage{psfrag}
\usepackage{bbm}
\usepackage{bm}
\usepackage{titlesec}
\usepackage{dsfont}
\usepackage{feynmp}
\usepackage{slashed}
\usepackage{multirow}
\usepackage[tight]{subfigure}

\usepackage[papersize={8.5in,11in}]{geometry}

\usepackage{color}
\definecolor{darkblue}{rgb}{0.,0.,0.4}
\definecolor{darkred}{rgb}{0.5,0.,0.}
\definecolor{BlueViolet}{RGB}{138,43,226}
\definecolor{SkyBlue}{RGB}{30,144,255}
\definecolor{DarkGreen}{RGB}{0,100,0}
\usepackage[pdftex,colorlinks=true,linkcolor=darkblue,citecolor=blue,urlcolor=darkred]{hyperref}

\usepackage{physics}

\def \nn{\nonumber \\}

\begin{document}

\title{Effects of time-periodic drive in the linear response for planar-Hall set-ups with Weyl and multi-Weyl semimetals}

\author{Ipsita Mandal}

\email{ipsita.mandal@snu.edu.in}

\affiliation{Department of Physics, Shiv Nadar Institution of Eminence (SNIoE), Gautam Buddha Nagar, Uttar Pradesh 201314, India}

\begin{abstract}
We investigate the influence of a time-periodic drive on three-dimensional Weyl and multi-Weyl
semimetals in planar-Hall/planar-thermal-Hall set-ups. The drive is modelled here by circularly-polarized electromagnetic fields, whose effects are incorporated by a combination of the Floquet theorem and the van Vleck perturbation theory, applicable in the high-frequency limit. We evaluate the longitudinal and in-plane transverse components of the linear-response coefficients using the semiclassical Boltzmann formalism, thereby demonstrating the explicit analytical expressions of the conductivity for large frequencies. Our results corroborate the fact that the topological charges of the corresponding semimetals etch their trademark signatures in these transport properties, which can be detected in appropriate experiments.
\end{abstract}


\maketitle

\tableofcontents

\section{Introduction}

\begin{figure}[t]
	\centering
\subfigure[]{\includegraphics[width=0.67 \textwidth]{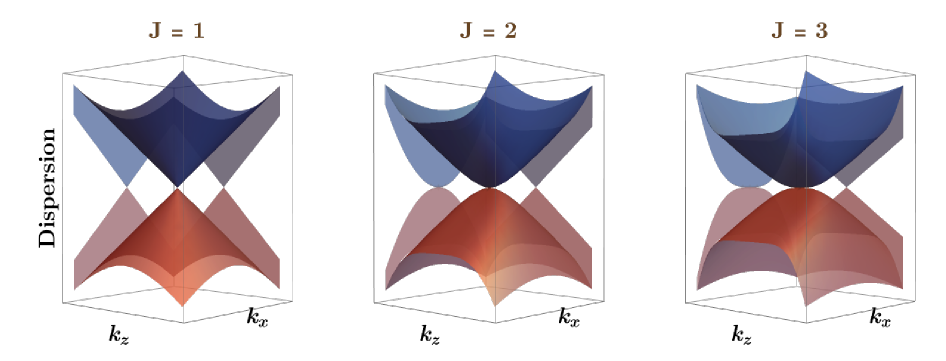}} \quad
\subfigure[]{\includegraphics[width=0.3\textwidth]{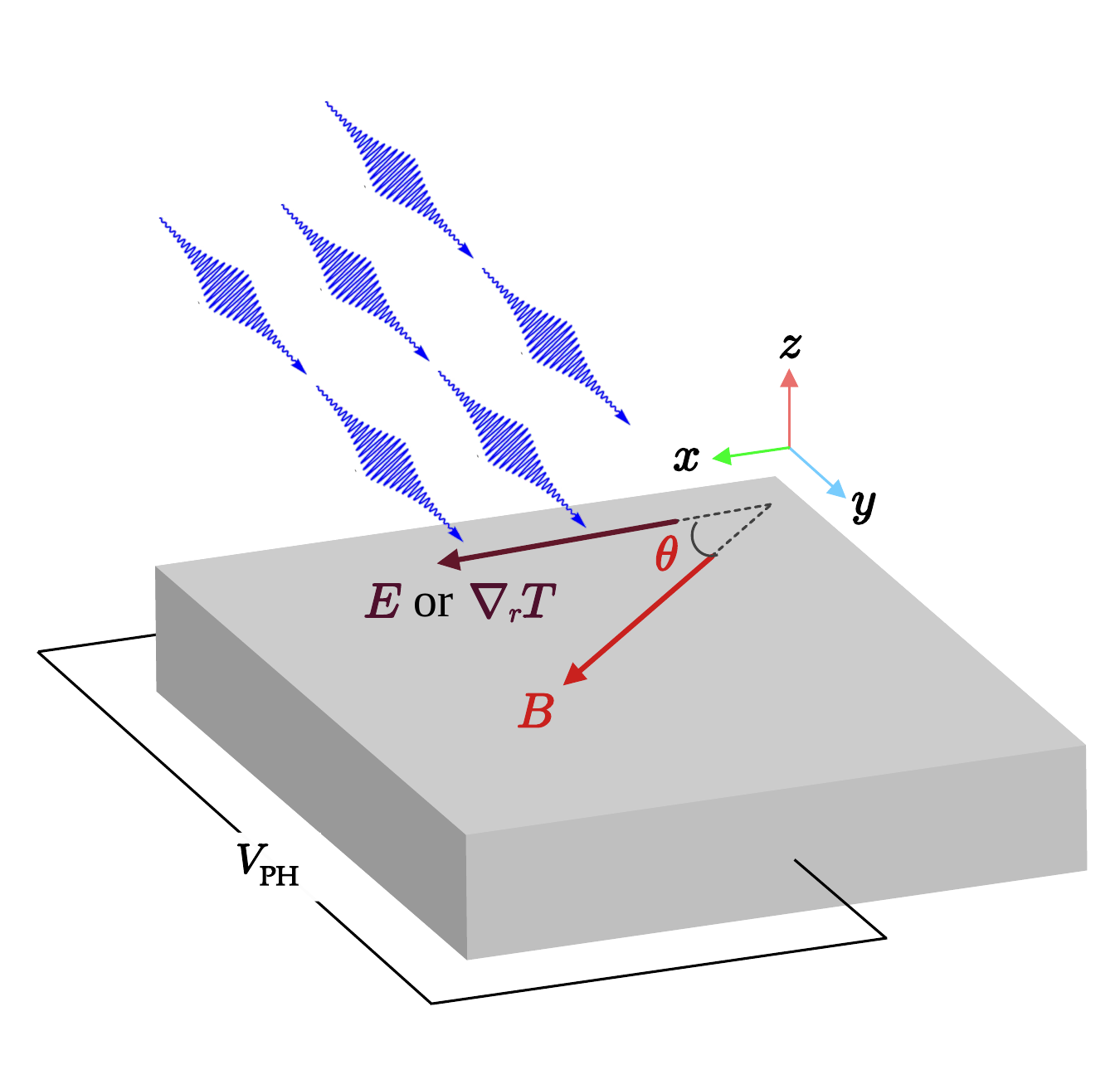}}
\caption{\label{figsetup}
(a) Dispersion characteristics of a single node in a Weyl, double-Weyl, and triple-Weyl semimetal, respectively,plotted against the $k_z k_x $-plane. The double(triple)-Weyl node shows an anisotropic hybrid dispersion with a quadratic(cubic)-in-momentum dependence along the $k_x$-direction. In order to pinpoint the direction-dependent features, the projections of the dispersion along the respective momentum axes are also shown. (b) Schematics showing the planar-Hall (or planar-thermal Hall) experimental configuration, where the sample is subjected to a uniform electric field $ E\, {\boldsymbol{\hat x}} $ (and/or a temperature gradient $\partial_x T\, {\boldsymbol{\hat x}}$) along the $x$-axis. An external magnetic field $\boldsymbol B $ is applied as well, which makes an angle $\theta $ with the $x$-axis. The blue wave-packets represent a time-periodic drive implemented by shining circularly-polarized light.}
\end{figure}

In contemporary research involving solid-state materials, there has been an upsurge in the explorations of systems exhibiting band-crossing points in the Brillouin zone, where the densities-of-states go to zero. These include the celebrated Weyl semimetals (WSMs)\cite{hasan} and its cousins, the multi-Weyl semimetals (mWSMs) \cite{bernevig12}. These are three-dimensional (3d) semimetals hosting nodal points, which also harbour nontrivial topological properties in their bandstructures. The topology of the 3d manifold formed by the Brillouin zone is responsible for giving rise to various novel properties. Fermi arcs, chiral anomaly, planar-Hall effects exemplify some such exotic characteristics. The nodal points behave as sinks and sources of the Berry flux --- i.e., they act like the monopoles of the Berry-curvature (BC) vector field, which arises from the Berry phase. Since the total topological charge over the entire Brillouin zone must vanish, these nodes must come in pairs, each pair carrying positive and negative topological charges of equal magnitude. This also follows from the Nielsen–Ninomiya theorem \cite{NIELSEN}. Mathematically, these monopoles (or topological charges) are equivalent to the Chern numbers. The sign of the monopole charge is often referred to as the chirality of the corresponding node. While WSMs have a linear and isotropic dispersion and their band-crossing points harbour Chern numbers equalling $\pm 1$, mWSMs exhibit anisotropic dispersions, which turn out to be a hybrid of linear and nonlinear (in momentum) [cf. Fig.~\ref{figsetup}(a)]. The mWSMs, additionally, harbour nodes with Chern numbers $\pm 2$ (double-Weyl) or $\pm3$ (triple-Weyl). It can be proved mathematically that the magnitude of the Chern number in mWSMs is bounded by $3$, by using symmetry arguments for crystalline structures \cite{Xu:2011, Nagaosa_2014, bernevig12}. Due to the nonzero topological charges of these systems, novel optical and transport properties, such as circular photogalvanic effect \cite{Moore_Joel}, circular dichroism \cite{sajid_cd}, negative magnetoresistance \cite{Lv_2021, Huang_et.al}, and planar-Hall effect (PHE) \cite{burkov}, magneto-optical conductivity~\cite{ashby,magneto-double-weyl,marcus-emil, ips-shivam} can emerge. 

There has been unprecedented advancement in the experimental fronts as well, where WSMs have been realized experimentally \cite{ding,ding2,Huang_et.al, Xu_et.al, Su_Yang_et.al} in compounds like TaA, NbA, and TaP. These materials have been reported to have topological charges equal to $\pm 1$. Compounds like $\text{HgCr}_{2}\text{Se}_{4}$ and $\text{SrSi}_{2} $ have been predicted to harbour double-Weyl nodes \cite{Xu:2011, bernevig12, Huang1180}. DFT calculations have found that nodal points in compounds of the form $\text{A(MoX)}_{3}$ (where A = Na, K, Rb, In, Tl, and X=S, Se, Te) have Chern numbers $\pm 3$ \cite{Liu_2017}.
Dynamical/nonequilibrium topological semimetallic phases can also be designed by Floquet engineering \cite{sentef, PhysRevE.93.022209,PhysRevB.103.094309,Umer_2021}.

When a conductor is placed in a magnetic field $\boldsymbol{B}$, such that it has a nonzero component perpendicular to the electric field $\boldsymbol{E}$ (which has been applied across the conductor), a current is generated perpendicular to the $\boldsymbol{E}$-$\boldsymbol B$ plane. This current is usually referred to as the Hall current, and the phenomenon is the well-known Hall effect. A generalization of this phenomenon is the PHE \cite{burkov}, when there is the emergence of a voltage difference perpendicular to an applied external $\boldsymbol{E}$, which is in the plane along which $\boldsymbol{E}$ and $\boldsymbol{B}$ lie [cf. Fig.~\ref{figsetup}(b)]. The planar-Hall conductivity, denoted by $\sigma_{xy}$ in this paper, is dependent on the angle between $\boldsymbol{E}$ and $\boldsymbol{B}$. In contrast with the canonical Hall conductivity, PHE does not require a nonzero component of $\boldsymbol{B}$ perpendicular to $\boldsymbol{E}$. While this novel phenomenon has been known to exist in ferromagnetic materials \cite{ferro,Phy172401,Friedland_2006,gupta,bowen}, it has also been found to emerge in semimetals when they harbour nontrivial topology in their Brillouin zone. In particular, WSMs/mWSMs possess a nontrivial BC, which sources the so-called chiral anomaly \cite{chiral_ABJ, hosur-review, son13_chiral, ips-internode}. We would like to emphasize that the planar-Hall effect survives in a configuration in which the conventional Hall effect vanishes (because $\boldsymbol E$, $\boldsymbol B $, and the induced transverse Hall voltage, all lie in the same plane).
Similar to the PHE, the planar-thermal-Hall effect [also referred to as the planar Nernst effect (PNE)] is the appearance of a voltage gradient perpendicular to an applied temperature ($T$) gradient $\nabla_{\boldsymbol r} T$, which is coplanar with an externally-applied magnetic field $\boldsymbol B$ [cf. Fig.~\ref{figsetup}(b)]. 

There have been extensive theoretical \cite{Nandy_2019,Nag_2020,ips-kush, banasri_floquet, ips-hermann-thermo} and experimental \cite{Huang_et.al, Zyuzin_2017} studies of the transport coefficients in these planar-Hall set-ups for various semimetals. Examples include longitudinal magneto-conductivity (LMC), planar-Hall conductivity (PHC), longitudinal thermoelectric coefficient (LTEC), transverse thermoelectric coefficient (TTEC)
(also known as the Peltier coefficient), and the components of the thermal conductivity. In this paper, we will compute these linear-response transport coefficients for WSMs and mWSMs, where such a semimetal is subjected to a time-periodic drive (for example, by shining circularly-polarized light with frequency $\omega$). 
We will use the semiclassical Boltzmann approach for calculating these properties.

A widely-used approach to analyze periodically-driven systems, where the time-independent Hamiltonian is perturbed with a periodic potential, is the application of the Floquet formalism
\cite{eckardt,  banasri_floquet, Nag_floquet_2020, rui-cong, Oka_takashi, ips-sandip, ips-sandip-sajid}. The approach relies on the fact that a particle can gain or lose energy in multiples of $\hbar \omega$ (quantum of a photon), where $\omega$ is the driving frequency. Since the time($t$)-dependent Hamiltonian $H$ satisfies $H(t + \mathcal{T}) = H(t)$, where $\mathcal{T} = 2\,\pi/ \omega$, we perform a Fourier transformation in the time domain. When $\omega$ is much larger than the typical energy-bandwidth of the system, we can combine the Floquet formalism with the van Vleck perturbation theory to obtain an effective perturbative potential of the following form~\cite{eckardt}:
\begin{align}
	V_{\text{eff}} &= 
	\sum \limits_{n=1}^{ \infty} 
\left( \frac{ \left[H_{-n} ,H_{n} \right ] }  {n \,\omega} 
	+ \frac{ \left [ \left [H_{-n}, H_0 \right ] , H_{n} \right ]  } {n^2\, \omega^2} 
	+ \cdots \right).
\end{align}
Here, $H_{\tilde n}$ denotes the ${\tilde n}^{\rm{th}}$ Fourier mode of the Hamiltonian.

The paper is organized as follows: In Sec.~\ref{secmodel}, we explain the forms of the low-energy effective Hamiltonians for the WSMs and mWSMs. There, we also chalk out the formalism employed to compute the transport coefficiants in the planar-Hall set-ups, coupled with the application of a time-periodic drive. In Sec.~\ref{sec_conduct}, we derive the explicit expressions for the in-plane components of the magneto-electric conductivity, dubbed as the LMC and the PHC. In Sec.~\ref{sec_dis_con}, we discuss the distinct characteristics observed from our results. Finally, we conclude with a summary and outlook in Sec.~\ref{secsum}. The appendix contains some details regarding the explanations of the intermediate steps. In all our expressions, we will be using the natural units, which amounts to setting the reduced Planck's constant ($\hbar $), the speed of light ($c$), and the Boltzmann constant ($k_B $) to unity. Additionally, electric charge has no units, with the magnitude of a single electronic charge measuring $e =1$. Nevertheless, we will retain the symbol $e$ in our linear-response tensors purely for the purpose of book-keeping.


\section{Model and formalism}
\label{secmodel}

The low-energy effective Hamiltonian in the vicinity of a single WSM/mWSM node, with topological charge $J$,
can be written as \cite{Xu:2011, bernevig12, Nagaosa_2014}
\begin{align}
\label{eq_multi1}
&	H_J \left( \boldsymbol{k} \right) = \boldsymbol{d}_{\boldsymbol{k}}\cdot \boldsymbol{\sigma}
=     \begin{bmatrix}  
v_z \, k_z &  \alpha_J\left (k_x-i\,k_y \right)^J \\ 
		\alpha_J \left (k_x+ i\,k_y \right )^J  &   -  \, v_z \, k_z
\end{bmatrix},
\text{ with } J \in (1,2,3)	, \nn
&  \boldsymbol{d}_{\boldsymbol{k}}
=  \left \lbrace \alpha_J \,k^J_\perp\cos \left( J\, \phi_{k}  \right), 
\,\alpha_J \,k^J_\perp
\sin \left( J\, \phi_{k}  \right),  \, v_z \,k_z \right \rbrace, \quad
k_\perp = \sqrt{k_x^2+k_y^2} \,, \quad 
\phi_k = \arctan \Big (\frac{k_y}{k_x} \Big ) \,,\quad
\alpha_J = \frac{v_\perp}{k_0^{J-1}}\,.
\end{align}
Here, $v_z$ and $v_\perp$ are the Fermi velocities in the $z$-direction and $xy$-plane, respectively, and $k_0$ is a system-dependent parameter with the dimension of momentum.
The vector $\boldsymbol{\sigma} $ comprises the three Pauli matrices (viz. $ \sigma_x$, $\sigma_y$, and $\sigma_z $), defining the pseudospin space. The value of $J=1$ represents the WSM, which has a linear and isotropic dispersion with $v_\perp = v_z$.
The energy eigenvalues are given by $(-1)^m \,\epsilon^{(0)}_{\boldsymbol{k}}$, where
\begin{align}
	\epsilon^{(0)}_{\boldsymbol{k}} 
	=   \sqrt{ \alpha^2_J \,k^{2 J}_\perp + v_z^2\, k^2_z} 
\;	 \text{ and } m \in (0,1)\,,
	\label{eq_multi2}
\end{align}
with the values ``$0$'' and ``$1$'' representing the conduction and valence bands, respectively. 
The hybrid nature of the dispersion for $J>1$ is depicted in Fig.~\ref{figsetup}(a).

The quasiparticle-velocity (or group-velocity) vectors, associated with the conduction and valence bands, are given by
$ (-1)^m \, {\boldsymbol v}^{(0)} $, where
\begin{align}
{\boldsymbol v}^{(0)} \equiv 
\left \lbrace v^{(0)}_x,\,v^{(0)}_y,  \, v^{(0)}_z \right \rbrace
= {\nabla}_{\boldsymbol k} \epsilon^{(0)}_{\boldsymbol{k}}
= \frac{1}  {\epsilon^{(0)}_{\boldsymbol{k}} }
	\left \lbrace J\,k_x \, \alpha_J^2\, k_\perp^{2 \,J-2 },\,
	J\,k_y \, \alpha_J^2 \,k_\perp^{2 \,J- 2 }, \,v_z^2 \,k_z \right \rbrace.
\label{eqvel}
\end{align}

The $i^{\rm{th}}$-component of the BC for the $m^{\textrm{th}}$ band is given by \cite{Xiao_2010}
\begin{align}
\boldsymbol \Omega^{(m)}_i (\boldsymbol k)
& = \frac{ (-1)^{m}  \, \epsilon_{i j l} }
	{4}\,
\frac{	\boldsymbol{d}_{\boldsymbol{k}} 	
	\cdot \left( \partial_{k_j} \boldsymbol{d}_{\boldsymbol{k}}
	\times 
\partial_{k_l} \boldsymbol{d}_{\boldsymbol{k}} \right)}
{|\boldsymbol d_{\boldsymbol{k}}|^3}	\,,
\text{ with } i,\,j ,\, l  \in \lbrace x,y,z \rbrace \nn
\Rightarrow  \boldsymbol \Omega^{(m)} (\boldsymbol{k} )
& = \frac{  (-1)^{m} \,
J \,v_z\, \alpha_J^2 \,k^{2J-2}_\perp }
	{2\,\left( \epsilon^{(0)}_{\boldsymbol{k}} \right)^3 } 
	\left \lbrace k_x, \,k_y, \,J \,k_z \right \rbrace.
\label{eqbcl0}
\end{align}
Clearly, the components of ${\boldsymbol \Omega}^m (\boldsymbol{k}) $ are  anisotropic for $J>1$. 
We have chosen the convention such that the chirality (and, hence, the Chern number) is positive for the conduction band, taking the explicit form of
$ {\boldsymbol \Omega}^{(0)} (\boldsymbol{k}) =  
\frac{ J \,v_z\, \alpha_J^2 \,k^{2J-2}_\perp }
	{2\,\left( \epsilon^{(0)}_{\boldsymbol{k}} \right)^3 } 
	\left \lbrace k_x, \,k_y, \,J \,k_z \right \rbrace.$
In the following, we will consider the case when the chemical potential $\mu$ (measured with respect to the nodal point) cuts the positive-energy band (i.e., $\mu > 0$).

\subsection{Semiclassical Boltzmann formalism}
\label{secboltz}

In this subsection, we will review the semiclassical Boltzmann formalism~\cite{mermin, arovas, ips-kush-review, ips_rahul_ph_strain}, which is used to calculate the transport coefficients in the set-ups shown in Fig.~\ref{figsetup}(b). In particular, we limit ourselves to the regime when the externally-applied magnetic field ($\boldsymbol B$) has a small-enough magnitude, leading to a small cyclotron frequency $\omega_c = e\,B/m^* $ (where $m^* $ is the effective mass with the magnitude $\sim 0.11 \, m_e$, with $m_e$ denoting the electron mass), which satisfies $ \omega_c \ll \mu$. This condition ensures that we can ignore the quantization of the dispersion into the form of Landau levels.
While the details of the step-by-step derivations of the linear-response tensors can be found in Refs.~\cite{ips-kush-review, ips_rahul_ph_strain} in the context of nodal-point semimetals, we summarize the salient points here. When we use the Boltzmann equations to describe the evolution of the distribution function $f(\boldsymbol{k}, \boldsymbol{r},t)$ of fermionic quasiparticles, we start with the form given by
\begin{align} 
	\left( \frac{\partial}{\partial t} + \, \dot{\boldsymbol r}\, . {\nabla}_{\boldsymbol r} + \, \dot{\boldsymbol k} \, . {\nabla}_{\boldsymbol k} \right) f(\boldsymbol k, \boldsymbol r,t) 
	= I_{coll} [{f(\boldsymbol k,  \boldsymbol r ,t)} ]\,, 
\end{align}
which results from the Liouville’s equation in the presence of scattering events. 
The term on the right-hand side (viz. $ I_{coll} [{f(\boldsymbol k,  \boldsymbol r ,t)} ]$) represents the collision integral, which arises due to scattering of electrons (e.g., scattering from lattice or from impurities). In the relaxation-time approximation, $I_{coll} [{f(\boldsymbol k,  \boldsymbol r ,t)} ]$ is approximated as $ \delta f /
 \tau $, where (1) $\delta f = f - f_0 $, (2) $\tau$ is known as the relaxation time, and (3) $f_0$ is the equilibrium value of $f$ (i.e., in the absence of any externally-applied fields), captured by the Fermi-Dirac distribution function. Physically, $\tau$ is a phenomenological estimate of the average time between successive collisions of the quasiparticles. For the fermionic quasiparticles occupying a band with the dispersion represented as $\varepsilon (\boldsymbol k)$, $ f_0= f_0 ( \varepsilon (\boldsymbol k) , \mu, T) \equiv \left[ 1 + e^{\frac{\varepsilon (\boldsymbol k) -\mu } {T}}\right ]^{-1}$.

 Here, we are interested in looking for steady-state solutions, for which $f(\boldsymbol{k}, \boldsymbol{r},t)$ is time-independent, implying $ f(\boldsymbol{k}, \boldsymbol{r},t)
= f(\boldsymbol{k}, \boldsymbol{r})$. Under these circumstance, Eq.~\eqref{boltz_equ} reduces to
\begin{align}
\label{boltz_equ} 
	\left( \dot{\boldsymbol{r}} \cdot {\nabla}_{\boldsymbol{r}} 
	+ \dot{\boldsymbol{k}} \cdot {\nabla}_{\boldsymbol{k}} \right)  
	f({\boldsymbol{k}},\boldsymbol{r})
	=  \frac{- \,\delta f({\boldsymbol{k}},\boldsymbol{r})   }   { \tau }\,.
\end{align}
For a system harbouring a nontrivial BC \cite{horvathy,son, ips_rahul_ph_strain, ips-kush-review}, which we denote by $ {\boldsymbol \Omega}_F$, under the effect of a magnetic field $\boldsymbol B $, it is necessary to introduce a phase-space factor defined by
\begin{align}
\mathcal{D} ({\boldsymbol k})
&= \left [1+ e \; \boldsymbol{B} \cdot 
{\boldsymbol \Omega}_F({\boldsymbol k})   \right]^{-1},
\end{align}
which ensures that the Liouville’s theorem continues to hold in equilibrium (i.e., in the absence of any external probe fields). More specifically, $ {\mathcal{D}} ({\boldsymbol k}) $ arises from the fact that ${\boldsymbol{\Omega}}_F({\boldsymbol k})$ modifies the phase-space volume element as $d^3\boldsymbol k\, d^3\boldsymbol x \rightarrow d^3\boldsymbol k\, d^3\boldsymbol x\,\mathcal{D}^{-1}
({\boldsymbol k}) $. After incorporating this correction, the final forms of the semiclassical transport equations for a system with BC turn out to be 
\begin{align}
	\dot{\boldsymbol{r}} & = \mathcal{D} 
	\left[ \boldsymbol{v} + e \left ( \boldsymbol{E} \times {\boldsymbol \Omega}_F \right ) 
+ e \left (\boldsymbol{v} \cdot {\boldsymbol \Omega}_F \right ) \boldsymbol{B}  \right]\,,\quad
	\dot{\boldsymbol{k}} = -
	\mathcal{D} 
	\left[ e \,\boldsymbol{E} + e  \left( \boldsymbol{v} \times \boldsymbol{B} \right) 
	+ e^2  \left(\boldsymbol{E} \cdot \boldsymbol{B} \right ) {\boldsymbol \Omega}_F  \right ]	.
\label{eq_rk}
\end{align}
Here, $\boldsymbol v$ denotes the quasiparticles' group-velocity. 
In the semiclassical approach, the quasiparticles in the vicinity of a nodal point are described classically by wavepackets centered at $( \mathbf r,\,  \mathbf k)$. The effect of ``quantumness'' of the system is captured by additional terms in the applicable Hamilton's equations of the wavepackets, which represent topological properties like the BC \cite{Xiao_2010, pronin2023linear}. This is because, when we measure the electronic properties of materials (e.g., via transport properties), we deal with wavepackets rather than single Bloch waves. The quasiclassical approximation is thus often relevant for such measurements. As the wavepackets propagate through the material, they disperse (i.e., become broader) both in the real and reciprocal spaces. The evolution of the components of the wavepackets can be described by an appropriate Schrödinger-like equation. Now, both classical and quantum physics tells us that the propagation of the wavepackets in real space is described by their group-velocity. It can be shown relatively easily that the presence of Berry phase necessarily leads to an additional group-velocity, which is commonly called anomalous velocity [cf. Eq.~\eqref{eq_rk}].

Incorporating all the ingredients described above, Eq.~\eqref{boltz_equ} reduces to
\begin{align}
\label{eqkin33}
& \mathcal{D} \,
\left [ \left \lbrace  \boldsymbol{v}
+ e \,{\boldsymbol E} \cross \boldsymbol{\Omega}_F
+ e \left(  {\boldsymbol \Omega}_F \cdot \boldsymbol{v} \right)
	\boldsymbol{B} \right \rbrace 
	\cdot \nabla_{\boldsymbol r} 
-  e \left(  {\boldsymbol E} + \boldsymbol{v}\cross {\boldsymbol B}	\right) 
	\cdot \nabla_{\boldsymbol k} 
- e^2 \left ( {\boldsymbol E}  \cdot {\boldsymbol B} \right )
\,	{\boldsymbol \Omega}_F  \cdot \nabla_{\boldsymbol k} \right ] f 
= \frac{  - \,\delta  f }   { \tau } \,.
\end{align}
Next, one assumes that the probe fields [viz. $\boldsymbol E $ and/or $\nabla_{\boldsymbol r} T$] and the resulting $\delta f $ are of the same order of smallness. To the leading order in this ``smallness parameter'', the so-called \textit{linearized Boltzmann equation} is obtained as
\begin{align}
\label{eqlin}
& -e \left [
\left \lbrace {\boldsymbol{v}} 
+ e \left(
{\boldsymbol \Omega}_F \cdot {\boldsymbol{v}}   \right)  \boldsymbol B \right \rbrace
\cdot  \boldsymbol E \right] 
\frac{\partial  f_0 (\varepsilon , \mu, T) } {\partial \varepsilon }
+   e \, {\boldsymbol B} \cdot
	\left( {\boldsymbol v}  \cross \nabla_{\boldsymbol k}\right)  \, \delta f
  = -\frac{\delta f} 
{  \tau  \, \mathcal D  } \,.
\end{align}
Now, the charge- and thermal-current densities are given by \cite{mermin, arovas, ips-kush-review}
\begin{align}
	J^{i} & =  \sigma_{ij} \,E^{j} 
	+ \alpha_{ij}\,(-{\nabla}_{r_j} T)\, ,
\quad	J^Q_{i}  =  T\, \alpha_{ij} \,E^{j}
	+ \ell_{ij} \, (-{\nabla}_{r_j} T)\,,
	\label{e03}
\end{align}
respectively. Here, $ \sigma_{ij} $ and $ \alpha_{ij}$ define the components of the electric-conductivity tensor and the thermoelectric coefficients, respectively. The third one, namely $\ell $, is the tensor relating the thermal-current density to the temperature gradient at a vanishing electric field. Since $\ell  $ contributes to the magnetothermal conductivity tensor, we will loosely refer to $\ell_{ij}  $ themselves as the magnetothermal coefficients.

For investigating the response in the planar-Hall and planar-thermal-Hall set-ups, we consider the application of an external electric field $\boldsymbol{E}$ and/or temperature gradient along the $x$-axis [viz. $\boldsymbol{E}= E \, \boldsymbol {{\hat x}}$ and $\nabla_{\boldsymbol r} = \partial_x T\, \boldsymbol {{\hat x}}$], accompanied by an external magnetic field $\boldsymbol{B}$ along the $xy$-plane [viz., $\boldsymbol{B} = B_x\, \boldsymbol {{\hat x}} + B_y\, \boldsymbol {{\hat y}} $]. We will denote the angle made by $\boldsymbol{B}$ with respect to the $x$-axis as $\theta$, such that $B_x = B  \cos \theta $ and $ B_y = B  \sin \theta $. Now the charge- and thermal-current densities are given by
\begin{align}
\label{eqcur}
{\boldsymbol J} & =   -\, e \,   \int
\frac{ d^3 \boldsymbol k}{(2\, \pi)^3 } \,\mathcal{D}^{-1}    \, \dot{\boldsymbol r}
\,  f( \boldsymbol r , \boldsymbol k) \text{ and }
\text{ and }
 {\boldsymbol J}^Q  =   \int
\frac{ d^3 \mathbf k}{(2\, \pi)^3 } 
\, \mathcal{D}^{-1}   \, \dot{\bf r} \,
\left( \varepsilon - \mu \right)  f ( \bf r , \bf k)\,,
\end{align}
respectively, where $f$ is determined as the solution of Eq.~\eqref{eqlin} (see, for example, the appendix of Ref.~\cite{ips-shreya}). In particular, we will focus on the behaviour of the in-plane components (i.e., the $xx$- and $yx$-components) of the linear-response tensors, which are obtained by evaluating \cite{ips-ruiz, ips-rsw-ph, ips-shreya}
\begin{align}
\label{eqsigma}
&  \sigma_{ ij } \approx -\, e^2 \, \tau
\int \frac{ d^3 \boldsymbol k}{(2\, \pi)^3 } \, {\mathcal D}
\left [ v_i  + e\, B_i \left( 
{\boldsymbol{v}} \cdot \boldsymbol \Omega_F \right)
\right ]
\left [ v_j  + e\,  B_j 
\left( {\boldsymbol{v}} \cdot \boldsymbol \Omega_F \right)
\right ]
\, \frac{\partial  f_0 } {\partial  \varepsilon} \,,\nn
&  \alpha_{ ij } \approx e \, \tau
\int \frac{ d^3 \boldsymbol k} {(2\, \pi)^3 } \, \mathcal D
\left [ v_i  + e\, B_i \left( 
{\boldsymbol v} \cdot \boldsymbol \Omega_F \right)
\right ]
\left [ v_j  + e\,  B_j  \left( 
{\boldsymbol{v}}_s \cdot \boldsymbol \Omega_F \right)
\right ]
\, \frac{\left( \varepsilon - \mu \right) } {T}
\, \frac{\partial  f_0 } {\partial  \varepsilon} \,,\nn
&  \ell_{ ij } \approx -\, \tau
\int \frac{ d^3 \boldsymbol k} {(2\, \pi)^3 } \, \mathcal D
\left [ v_i  + e\, B_i \left( 
{\boldsymbol v} \cdot \boldsymbol \Omega_F \right)
\right ]
\left [ v_j  + e\,  B_j  \left( 
{\boldsymbol{v}}_s \cdot \boldsymbol \Omega_F \right)
\right ]
\, \frac{\left( \varepsilon - \mu \right)^2 } {T}
\, \frac{\partial  f_0 } {\partial  \varepsilon} \,.
\end{align}
We have used an ``approximately equal to'' ($\approx$) sign because we have ignored the contribution from the \textit{Lorentz-force} factors arising from external magnetic field [i.e., from the part $e \, {\boldsymbol B} \cdot
	\left( {\boldsymbol v}  \cross \nabla_{\boldsymbol k}\right)  \, \delta f$ in Eq.~\eqref{e03}]. This is justified because these corrections go to zero at leading order in the Lorentz-force operator, $\check L \equiv (\boldsymbol v \cross \boldsymbol{B}) \cdot \nabla_{\boldsymbol{k}}$, when we consider the in-plane components of the response tensors \cite{ips-rsw-ph, ips-spin1-ph,ips_tilted_dirac}.

A few important observations and comments are in order. We would like to point out that the phase-space factor of $\mathcal D$ in the expression for $ \alpha_{ ij }$ was \textbf{missed in Ref.~\cite{Nag_2020}} [cf. Eqs. (17) and (18) therein]. This, in some way, affected our computations in Ref.~\cite{ips-serena} (since we started with the erroneous expression of Ref.~\cite{Nag_2020}), which we have self-retracted after realizing the mistake.

\subsection{Time-periodic drive with a high frequency}

On top of the applied electric and magnetic fields ($ \boldsymbol{E}$ and $\boldsymbol{B}$), which are assumed to be static and uniform, we subject the system to a time-periodic optical drive with a high value of frequency, $\omega$. While the corresponding electric-field vector for the light waves can be represented as ${\boldsymbol{\mathcal E}}(t)
= E_0\left [-\cos (\omega  t) \, \boldsymbol {{\hat x}} +
 \sin (\omega \, t) \, \boldsymbol {{\hat y}} \right  ] $, a vector potential for the associated magnetic field can be written as $ {\boldsymbol A}(t) = \frac{E_0}{\omega} 
\left [ \sin (\omega \, t) \, \boldsymbol {{\hat x}} +  \cos(\omega \, t)  \, \boldsymbol {{\hat y }}\right ]$ upon using the Landau gauge. The effect of the circularly-polarized electromagnetic field on the Hamiltonian $H_J$ can be obtained via the Peierls substitution, $\boldsymbol k \rightarrow \boldsymbol k - e \,\boldsymbol  A $.
Defining $A_0= \frac{e \,E_0} {\omega}$, the gauge-dependent momentum components are found via $k_x \rightarrow k_x^\prime =
k_x -  A_0 \sin (\omega \, t)$,
$k_y  \rightarrow k_y ^\prime = k_y  -  A_0 \cos (\omega \, t)$, and  $k_z \rightarrow k_z^\prime = k_z $. Using the binomial expansion, we obtain $(k'_x  \, \pm \, i \, k'_y)^J
= \sum \limits_{m=0}^J (k_\perp \, e^{\pm i\,\phi_{k} })^{J-m} 
\,(- A_0)^m\, e^{\pm i \,m \,(\frac{\pi}{2}-\omega \, t )} ~^J C_m$,
where $~^J C_m = \frac{J!}{(J-m)! \,m!}$ 
represents the combinatorial factor. With these ingredients, the resulting time-dependent Hamiltonian takes the form of
\begin{align}
	\tilde H_J({\boldsymbol k},{\boldsymbol A}) & =
 \left [ 	\alpha_J \left (k'_x - i \,k'_y \right )^J \sigma_+
 + \alpha_J \left (k'_x + i\,k'_y \right)^J  \sigma_-
+   v_z \,k_z \, \sigma_z \right ] , \text{ where }
\quad \sigma_\pm =\frac{\sigma_x \pm i\,\sigma_y} {2} \,.
	\label{h1_time}
\end{align}
We consider the limit where Floquet theorem can be applied, thus extracting the leading-order-correction terms from van Vleck's high-frequency expansion . In this limit, one can describe the dynamics of the driven system, with time period $ \mathcal{T}=\frac{2 \, \pi}{\omega}$, in terms of an effective Floquet Hamiltonian defined as~\cite{eckardt}
\begin{align}
\label{eqh4}
H^F_J({\boldsymbol k})
= H_J({\boldsymbol k}) + V({\boldsymbol k}) + \mathcal{O} (\omega^{-2}), \text{ where }
 V ({\boldsymbol k}) =
\sum \limits_{p=1}^\infty \frac{[V_{-p},V_p] }  {p\, \omega} \text{ and }
V_p = \frac{1}{\mathcal{T}} \int_0^\mathcal{T}  dt\,
 \tilde H_J({\boldsymbol k}, {\boldsymbol A}) \, e^{i \,p\, \omega \, t} \,.
\end{align} 
It represents the driving term being treated perturbatively. 
The above approximation for $ H^F_J({\boldsymbol k})$ results from the fact that, while deriving the explicit analytical expressions for the linear response, we are going to restrict ourselves to only the leading-order pertubative term.

The explicit form of $V_p $ in Eq.~\eqref{eqh4} is obtained as
\begin{align}
	\label{v1}
	V_p =  \alpha_J \sum \limits_{m=1}^J  (k_\perp)^{J-m}(-A_0)^m ~^J C_m 
\begin{bmatrix} 
0 & e^{-i [(J-m) \phi_{k} +m\,\frac{\pi}{2}]} \,\delta_{p,-m} \\  
e^{i [(J-m) \phi_{k} +m\frac{\pi}{2}]} \,\delta_{p,m} & 0 
	\end{bmatrix},
\end{align} 
resulting in
\begin{align}
\label{h5}
H^F_J ({\boldsymbol k} ) &  \simeq
	{\boldsymbol d}_{\boldsymbol k} \cdot {\boldsymbol{\sigma}} 
	+ 
\lambda \, \frac{\alpha_J^2}{\omega} 
\sum \limits_{p=1}^J \frac{\left( \,^J C_p\, A_0^p \right)^2\, 
	k_\perp^{2J-2p }}{p} \;\sigma_z  
\equiv  \boldsymbol{\tilde d}_{\boldsymbol {k}}
\cdot {\boldsymbol \sigma} \,,\nn
\text{where } 
\boldsymbol{\tilde d}_{\boldsymbol {k}} & =
\left \lbrace  \alpha_J \, k^J_\perp\cos \left( J\, \phi_{k} \right ) , 
\,  \alpha_J \, k^J_\perp \sin \left( J\, \phi_{k} \right), \, 
v_z\, k_z + \lambda \,\frac{T_{\boldsymbol k}} {\omega} \right \rbrace \text{ and }
T_{\boldsymbol k}  =
\alpha_J^2 \, k_\perp^{2\,J}
\sum \limits_{p=1}^J 
\frac{(^J C_p \,A_0^p)^2} {p \, k_\perp^{2\,p}}  \,.
\end{align}
We have incorporated a parameter $\lambda $ infront of the perturbation term merely for the sake of book-keeping, so that we can keep track of the order to which we obtain the perturbative expansions. In the end of the calculations, we will set it to one.

Let us estimate the range of values for $ E_0$ so that we are in the perturbative regime. 
Considering the fact that the dominant contributions from quasiparticle-excitations to any transport property arise from the momentum regions around the Fermi surface, the leading contributions to the integrals [shown in Eq.~\eqref{eqsigma}] come from the region where $ v_z\, k_z  \sim \mu$ and $\alpha_J \, k_\perp^{J} \sim \mu$. From the first and the last terms in the series in $T_{\boldsymbol k}$, we have two conditions:
\begin{align}
E_0^2 \ll  \frac{\omega ^3 }{e^2 J \mu ^2}  \left(\frac {\mu } {\alpha_J}  \right)^{2/J}
\text{ and } \quad
E_0^2 \ll
\frac{\omega ^2} {e^2}
 \left(\frac {J \,\omega } {\alpha_J^2} \right)^{1/J} \,.
\end{align}
If we consider the representative values in Table~\ref{table_quantites} (and with $\omega > \mu$), we find that the stricter bound is always provided by the second condition, from which we conclude that we must have $E_0^2 \ll
\frac{\omega ^2} {e^2}
 \left(\frac {J \,\omega } {\alpha_J^2} \right)^{1/J}$, representing the bounds on the values of $E_0 $ within which our analytical approximations are valid.

The eigenvalues of this effective Hamiltonian, which we call quasi-energies, are given by $ (-1)^m \,\epsilon_{\boldsymbol{k}}$, with
\begin{align}
\epsilon_{\boldsymbol{k}}  =   \sqrt{ \alpha^2_J \,k^{2 J}_\perp 
+ \left (v_z\, k_z + \lambda\,\frac{T_{\boldsymbol k}} {\omega} \right )^2 }\,.
	\label{eff_eng}
\end{align}
As before, $m =0$ ($ m=1$) refers to the conduction (valence) band. The group-velocity vectors are also modified to $ (-1)^m \, {\boldsymbol v} $, where
\begin{align}
{\boldsymbol v}  = {\nabla}_{\boldsymbol k} \epsilon_{\boldsymbol{k}}
= \frac{1}  {\epsilon_{\boldsymbol{k}} }
	\left \lbrace J\,k_x \, \alpha_J^2\, k_\perp^{2 \,J-2 },\,
	J\,k_y \, \alpha_J^2 \,k_\perp^{2 \,J- 2 }, \,v_z^2 \,k_z \right \rbrace .
\end{align}
These modified values, compared to the static/undriven system represented by Eq.~\eqref{eff_eng}, will affect the behaviour of the transport coefficients.

Working with $\tilde H_J({\boldsymbol k},{\boldsymbol A})$, the BC of the conduction band of the effective system is given by
\begin{align}
	{\boldsymbol \Omega}^{(0)}_F({\boldsymbol{k}}) & =
	\frac{ \alpha_J^2 \,k^{2J-2}_\perp\,J }
{ 2\,\epsilon_{\boldsymbol{k}}^3}
	\left \lbrace  v_z\,k_x, \,
v_z\, k_y, 
\,J\,v_z\, k_z + \lambda \,
  \frac{ J\, T_{\boldsymbol k}  
  - \Upsilon_{\boldsymbol k} } {\omega} \right \rbrace ,
\text{ where } \Upsilon_{\boldsymbol k} 
=  2\,\alpha_J^2 \,  k_\perp^{2 \, J}
\sum \limits_{p=1}^J 
\left (J-p \right) 
\frac{(^J C_p \,A_0^p)^2} {p \, k_\perp^{2\,p}} \,.	
\label{eq_bcl}
\end{align}
Comparing this with Eq.~\eqref{eqbcl0}, we note that the $z$-component of the BC has changed. This is expected to produce a discernible effect via terms involving
\begin{align}
\label{eq_vals}
&   \boldsymbol{v}\cdot {\boldsymbol{\Omega}}_F^{(0)}  = 
\frac{J \,k_\perp^{2 \,J - 2} \,v_z\, \alpha_J^2} 
{2 \,\epsilon_{\bf k}^4}
\left [ J \left (\epsilon^{(0)}_{\bf k} \right)^2
+ \lambda \,
\frac{k_z \left(J \,T_{\bf k}-U_{\bf k}\right) v_z}{\omega } \right ] 
\text{ and }
  \boldsymbol{B} \cdot {\boldsymbol \Omega}^{(0)}_F
=
\frac{J \,k_\perp^{2\, J-2 } \,v_z \,\alpha_J^2
\left( B_x \,k_x  + B_y \,k_y   \right) }
{2 \,\epsilon_{\bf k}^3}\,.
\end{align}
These appear in the expressions for the integrands determining the transport coefficients \cite{arovas, ips-kush-review, ips_rahul_ph_strain}.

\begin{table}[]
	\begin{tabular}{|l|l|l|}
		\hline
Parameter &   SI Units &   Natural Units  \\ \hline
$v_z$ from Ref.~\cite{Watzman_2018} & $15 \times10^5 $ m~s$^{-1} $ & $0.005$  \\ \hline
$\tau$ from Ref.~\cite{Watzman_2018} & $~10^{-13} \, s $ & $151.72 $ eV$^{-1}$  \\ \hline
$E_0$ from Ref.~\cite{YH_wang_2013} & $2.5\times 10^{7} $ V~m$^{-1}$ & $ 57.75 $ eV$^2 $  \\ \hline
$T$ from Ref.~\cite{Nag_2020}
& $\sim 10 $ -- $100 $ K & $ \sim 8.617 \times 10^{-4} $ -- $ 8.617 \times 10^{-3} $ eV \\ \hline
$B$ from Ref.~\cite{Nandy_2017} &  $0$ -- $ 5 $ Tesla & $0$ -- $ 1000 $ eV$^2$  \\ \hline
$\omega$ from Ref.~\cite{Nag_floquet_2020} & 
$\sim 15 \times 10^{14} $ -- $\sim 15 \times 10^{15} $ Hz & $\sim 1$ -- $\sim 10 $ eV  \\ \hline
$\mu$ from Refs.~\cite{Nag_2020, Nag_floquet_2020} & $1.6 \times 10^{-20} $ J & $0.1$ eV \\ \hline
\end{tabular}
\caption{\label{table_quantites}
The values of typical parameter values applicable for WSMs/mWSMs are tabulated here.
Since $\alpha_J =  {v_\perp}/ {k_0^{J-1}}$,
$\alpha_{1}=v_\perp$,
$\alpha_{2}= {v_\perp} / {k_0}$, this implies that
$ \alpha_{3} = {v_\perp} / {k_0^2} = {\alpha_{2}^2} / {\alpha_{1}} $.
In natural units, $\hbar=c=k_{B}= e = 1$, and $4 \, \pi \,\epsilon_0 = 137$. In our plots, we have used $v_\perp = v_z$ (from the table entry), $\alpha_2 = 0.00012 $ eV$^{-1}$ \cite{Nag_floquet_2020}, and $\alpha_{3}= 2.88 \times 10^{-6}$ eV$^{-2}$. For $J=2$ and $J=3$, $v_\perp$ has been set equal to $ v_z$ for the sake of simplicity, while for $J=1$, we have the constraint $v_\perp = v_z$ anyway (due to isotropy).
}
\end{table}

\section{Components of the conductivity tensors}
\label{sec_conduct}

In this section, we will evaluate the integrals shown in Eq.~\eqref{eqsigma}. Primarily, we will derive the expressions for the LMC and the PHC for the situation when the chemical potential cuts the positive-energy band. As for the LTEC and TTEC, we will invoke the Mott relations \cite{mermin} to outline their structures. The Mott relations are applicable in the limit $T \rightarrow 0 $ \cite{mermin, ips_rahul_ph_strain, ips-ruiz, ips-rsw-ph, ips-spin1-ph}, and they continue to hold in the presence of a nonzero BC (see, for example, Eq.~\cite{xiao06_berry}, where generic settings for deriving linear response have been considered). 
Here, we show the results which are correct up to order $\lambda^2$ in the perturbative expansion.

Due to the presence of the phase-space factor $\mathcal D $ in each integral, which is incorrigible in the presence of a nontrivial BC, we need to expand the $ B $-dependent terms upto a given order in $B$ in order to obtain closed-form analytical expressions. This requires the assumption that $B$ has a small magnitude, which is anyway an essential condition to justify neglecting the quantization of the dispersion into discrete Landau levels (thus allowing the application of the semiclassical Boltzmann formalism, as discussed in Sec.~\ref{secboltz}).
Here, we will limit ourselves to the expansion up to order $B^2$, because those are the lowest-order $B$-dependent nonzero terms that can appear in the in-plane response tensors.

\subsection{Magneto-electric conductivity: LMC and PHC}
\label{secsig}

Using Eq.~\eqref{eqsigma}, the expressions for the LMC and the PHC are given by
\begin{align}
\sigma_{xx} &\approx  - \,e^2 \,\tau
\int\frac{d^{3} \boldsymbol k} {(2 \, \pi)^{3}} \,\mathcal{D} 
	\left [ v_x + e\,B_x 
	\left ( \boldsymbol{v}\cdot {\boldsymbol{\Omega}}_F^{(0)} \right ) \right ]^2 
\,  \partial_{\epsilon_{\boldsymbol{k}}} f_0 (\epsilon_{\boldsymbol{k}}) 
\text{ and}	
\nn \sigma_{yx} & \approx  - \, e^2  \,\tau
\int\frac{d^{3} \boldsymbol k}{(2 \, \pi)^{3}} \,\mathcal{D}
\left [v_y +{e\,B_y }
	\left ( \boldsymbol{v}\cdot {\boldsymbol{\Omega}}_F^{(0)} \right )
	\right ]	
	\left [ v_x  + e\,B_x 
	\left ( \boldsymbol{v}\cdot {\boldsymbol{\Omega}}_F^{(0)} \right )
	\right ] \,  \partial_{\epsilon_{\boldsymbol{k}}} f_0 (\epsilon_{\boldsymbol{k}})\,,
	\label{eq_lmc}
\end{align}
respectively.

For the convenience of computations, $\sigma_{yx}$ and $\sigma_{xx}$ are divided up as
\begin{align}
\label{eq_lmc_break}
\sigma_{xx}= \sigma_{xx}^{(1)}+\sigma_{xx}^{(2)} + \sigma_{xx}^{(3)}\,,\quad
\sigma_{yx}= \sigma_{yx}^{(1)}+\sigma_{yx}^{(2)}+\sigma_{yx}^{(3)} + \sigma_{yx}^{(4)}\,,
\end{align} 
where 
\begin{align}
\label{s0}
\sigma^{(1)}_{xx} &=  e^2 \,\tau  
\int \frac{d^3\boldsymbol{k}}{(2\,\pi)^3}
\mathcal{D} \,   v_x^2 
\left [- \partial_{\epsilon_{\boldsymbol{k}}} f_0 (\epsilon_{\boldsymbol{k}}) \right ] ,\quad
\sigma^{(2)}_{xx} = { e^4\, \tau \, B_x^2 }
\int \frac{d^3 \boldsymbol{k}}{(2\,\pi)^3} 
\mathcal{D} \left ( \boldsymbol{v}  \cdot \boldsymbol{\Omega}_F^{(0)} \right )^2
\left [- \partial_{\epsilon_{\boldsymbol{k}}} f_0 (\epsilon_{\boldsymbol{k}}) \right ] ,\nn
\sigma^{(3)}_{xx} &=  2 \,e^3 \,\tau  \,B_x 
\int \frac{d^3\boldsymbol{k}} {(2\,\pi)^3} 
\mathcal{D} \,v_ x \left (  \boldsymbol{v} \cdot \boldsymbol{\Omega}_F^{(0)} \right )
\left [- \partial_{\epsilon_{\boldsymbol{k}}} f_0 (\epsilon_{\boldsymbol{k}}) \right ] ,
\end{align}
and
\begin{align}
\label{syx}
\sigma^{(1)}_{yx} &=  e^2 \,\tau
 \int \frac{d^3\boldsymbol{k}}{(2\,\pi)^3}
\mathcal{D} \,  v_y\, v_x 
\left [- \partial_{\epsilon_{\boldsymbol{k}}} f_0 (\epsilon_{\boldsymbol{k}}) \right ] ,\quad
\sigma^{(2)}_{yx} = { e^4\,\tau\,  B_y\, B_x }
\int \frac{d^3 \boldsymbol{k}}{(2\,\pi)^3} 
\mathcal{D} \left ( \boldsymbol{v}  \cdot \boldsymbol{\Omega}_F^{(0)} \right )^2
\left [- \partial_{\epsilon_{\boldsymbol{k}}} f_0 (\epsilon_{\boldsymbol{k}}) \right ] ,\nn
\sigma^{(3)}_{yx} &=  e^3\,\tau \,B_x  
\int \frac{d^3\boldsymbol{k}}{(2\,\pi)^3} 
\mathcal{D} \,v_ y \left (  \boldsymbol{v}  \cdot \boldsymbol{\Omega}_F^{(0)} \right )
\left [- \partial_{\epsilon_{\boldsymbol{k}}} f_0 (\epsilon_{\boldsymbol{k}}) \right ] ,\quad
\sigma^{(4)}_{yx} =  e^3 \, \tau \,B_y
	\int \frac{d^3\boldsymbol{k}}{(2\,\pi)^3} 
\mathcal{D} \,v_ x \left (  \boldsymbol{v} \cdot \boldsymbol{\Omega}_F^{(0)} \right )
\left [- \partial_{\epsilon_{\boldsymbol{k}}} f_0 (\epsilon_{\boldsymbol{k}}) \right ] .
\end{align}
We consider the zero-temperature limit ($T=0$) and, hence, 
$ - \,\partial_{\epsilon_{\boldsymbol{k}}} f_0 (\epsilon_{\boldsymbol{k}})
= \delta (\epsilon_{\boldsymbol{k}} -\mu )$. The finite-temperature results can be easily obtained
from the zero-temperature result using the relation \cite{mermin}
\begin{align}
\label{eqfiniteT}
\sigma_{ij} (\mu, T) = -\, \int_{-\infty}^\infty d\varepsilon \,
\sigma_{ij} (\varepsilon, T=0) \,
\partial_\varepsilon f_0(\varepsilon, \mu, T)\,.
\end{align}
We provide the elaborate and complete expressions for the in-plane components for all the three values of $J$, dividing them under three separate heads.

\subsubsection{$J=1$}

The three parts of the LMC turn out to be
\begin{align}
& \sigma_{xx}^{(1)} =
6 \,\pi ^2 \,e^2 \,\tau  \,v_z 
\left(\mu ^2-\frac{\alpha_1^4\, A_0^4} {5\, \omega ^2}\right)
+ 
\frac{e^4 \,\tau \,  \alpha_1^2\, v_z \left(3 B_x^2+B_y^2\right)}
{120\,   \pi ^2 \mu ^2}
 \left(1-\frac{3 \, \alpha_1^4 \,A_0^4} {7\, \mu ^2\, \omega ^2}\right)  , 
\nn &
\sigma_{xx}^{(2)} =
\frac{ e^4\,\tau \, \alpha_1^2 \,  B_x^2 \, v_z} {8\, \pi ^2 \, \mu ^2}
\left (
1-\frac{2 \,\alpha_1^4 \,A_0^4}   { 3 \, \mu ^2 \, \omega ^2}
\right ),\quad
\sigma_{xx}^{(3)} =
\frac{ - \,e^4 \, \tau \,\alpha_1^2  \, v_z \, B_x^2 } {60 \, \pi ^2 \, \mu ^2}
\left( 
5-\frac{3 \,\alpha_1^4 \,A_0^4} {\mu ^2\, \omega ^2}
\right ).
\end{align}
Similarly, the expressions for PHC are given by
\begin{align}
\sigma_{yx}^{(1)} = \frac{e^4\,\tau \,  v_z  \,\alpha_1^2  \, B_x  \, B_y}
{420 \,  \pi ^2\, \mu ^2}
\left( 
7-\frac{3 \,\, \alpha_1^4 \, A_0^4} {\mu ^2 \,\omega ^2}
\right) , \quad
\sigma_{yx}^{(2)} = \frac{B_y} {B_x} \, \sigma_{xx}^{(2)} \,,\quad
\sigma_{yx}^{(3)} = \sigma_{yx}^{(4)} = \frac{B_y} {B_x} \,\frac{ \sigma_{xx}^{(2)}} {2}\,.
\end{align}

Adding up all the parts, we obtain
\begin{align}
  \sigma_{xx}-  \sigma_{xx} \big \vert_{B=0} = 
\frac{e^4 \, \tau\, v_z \,\alpha_1^2 \, B^2 } {240\, \pi ^2\, \mu ^2}
\left[ 9 + 7 \cos (2 \theta )
-\frac{2 \, E_0^4 \,\alpha_1^4 
\left \lbrace 17 \cos (2 \theta )+20 \right \rbrace }
{7 \,\mu ^2\, \omega ^6} 
\right ]
\end{align}
and
\begin{align}
  \sigma_{yx} =
\frac{ 7\, e^4 \,\tau \, v_z\,\alpha_1^2 \,B^2 \sin (2 \theta )} 
{ 240\, \pi ^2\, \mu ^2}
\left(
1-\frac{ 34 \, E_0^4 \,\alpha_1^4} { 49\, \mu ^2 \, \omega ^6}
\right) .
\end{align}

\subsubsection{$J=2$}

The longitudinal and in-plane transverse components are obtained as follows:
\begin{align}
& \sigma_{xx}^{(1)} =
\frac{e^2 \, \tau \, \mu ^2 } {3 \,\pi ^2 \, v_z}
\left [1-
\frac{\alpha_2^2 \,A_0^4 \left(315 \,\pi \, \alpha_2\, A_0^2 \,\mu 
+ 28 \, \alpha_2^2 \, A_0^4 + 6144 \, \mu ^2\right)}
{560 \, \mu ^2 \, \omega ^2}
\right ]
+
\frac{5 \,e^4 \,\tau  \,v_z \, \alpha_2 \left(3 \,B_x^2 + B_y^2\right)}
{256 \, \pi \, \mu } \,
\left(1-\frac{\alpha_2^4 \,A_0^4}
{32\, \mu ^2\, \omega ^2}\right),\nn
& \sigma_{xx}^{(2)} =
\frac{ e^4 \,\tau  \,
\alpha_2 \, B_x^2 \, v_z} {4 \,\pi  \mu }
\left [1 + 
\frac{480 \,\pi \, \alpha_2^2 \,A_0^4 \,\mu + 224\, \alpha_2^3 \, A_0^6}
{15 \, \pi \, \mu \, \omega ^2}
\right  ], \nn &
\sigma_{xx}^{(3)} =
\frac{ - \,3 \, e^4\, \tau\, \alpha_2 \,v_z\,  B_x^2 }
{16 \, \pi \, \mu }
\left[1 +
\frac{ 4800 \,\pi \, \alpha_2^2 \,A_0^4\, \mu ^2
+ 2048 \, \alpha_2^3 \, A_0^6 \, \mu - 15 \, \pi \, \alpha_2^4 \, A_0^8}
{360\, \pi \, \mu ^2 \, \omega ^2}
\right ],
\end{align}
\begin{align}
& \sigma_{yx}^{(1)} =
\frac{5 \, e^4 \,\tau \, v_z\, \alpha_2} {128 \, \pi \, \mu }
\left(1-\frac{\alpha_2^4 \, A_0^8} {32\, \mu ^2\, \omega ^2}
\right),\quad
\sigma_{yx}^{(2)} = \frac{B_y} {B_x} \, \sigma_{xx}^{(2)} \,,\quad
\sigma_{yx}^{(3)} = \sigma_{yx}^{(4)} = \frac{B_y} {B_x} \,\frac{ \sigma_{xx}^{(2)}} {2}\,.
\end{align}

Adding up all the parts, we obtain
\begin{align}
  \sigma_{xx}-  \sigma_{xx} \big \vert_{B=0} = 
\frac{e^4 \,\tau \, v_z \,\alpha_2\, B^2 } {256\, \pi \, \mu }
\left [ 18 + 13 \cos (2 \theta ) +
\frac{1408 \,E_0^4 \,\alpha_2^2 \cos ^2 \theta} {\omega ^6}
+ \frac{2048 \, E_0^6 \,\alpha_2^3 \cos ^2 \theta } 
{3 \, \pi  \,\mu \, \omega^8} + 
\frac{ E_0^8 \,\alpha_2^4 \left \lbrace 22 + 27 \cos (2 \theta ) \right \rbrace }
{32 \, \mu ^2  \,\omega ^{10}}
\right ]
\end{align}
and
\begin{align}
  \sigma_{yx} =
\frac{13 \, e^4 \, \tau \,v_z \, \alpha_2 \,  B^2   \sin (2 \theta )} 
{256 \pi  \mu } 
\left( 1 +
\frac{704\, E_0^4\,\alpha _2^2}{13\, \omega ^6}
+ \frac{1024 \,E_0^6\, \alpha _2^3} {39\, \pi  \,\mu  \,\omega ^8}
+ \frac{27 \,E_0^8\, \alpha _2^4} {416 \, \mu ^2 \,\omega ^{10}}
\right ).
\end{align}

\subsubsection{$J=3$}

The LMC and the PHC are obtained as follows:
\begin{align}
& \sigma_{xx}^{(1)} =
\frac{e^2\, \tau\, \mu ^2  } {2 \, \pi ^2 \, v_z} 
\left [ 1 
-\frac{ 12 \,A_0^4 \, \Gamma
   \left(\frac{1}{6}\right) 
   \Gamma \left(\frac{1}{3}\right) \left(\frac{\alpha_3} {\mu }\right)^{ \frac{4} {3} } 
   \left(230 \alpha_3^2\, A_0^6 + 14553 \, \mu ^2\right)} 
   {21505 \, \sqrt{\pi } \,\omega^2 }
-\frac{\alpha_3^2 \,A_0^6 
 \left( 7 \,\alpha_3^2 \,A_0^6 +   17496 \, \mu ^2 \right)} 
{315 \, \mu^2 \,\omega^2 } 
-\frac{1750 \,\alpha_3^{ \frac{8} {3} } \,A_0^8 \, \Gamma \left(-\frac{4}{3}\right) 
\Gamma \left(\frac{5}{6}\right)} {247 \,\sqrt{\pi }\, \mu^{ \frac{2} {3} } \,\omega^2
} \right ]
\nn & \quad \qquad +
\frac{3200 \, e^4 \, \tau \,v_z \, \Gamma \left(-\frac{16}{3}\right) \,
\Gamma \left(\frac{5}{6}\right) 
\left(\frac{\alpha_3}
{\mu  }\right)^{ \frac{2} {3} }  \left(3 \,B_x^2 + B_y^2\right)} {171 \, \pi ^{ \frac{5} {2} }}
\, \Bigg[  1 +
\frac{741\, A_0^4 
\left \lbrace 
8575 \, \alpha_3^{ \frac{8} {3} } A_0^4 
\,\Gamma \left(\frac{1}{6} \right) 
\Gamma \left(\frac{1}{3}\right)
- 1632816\,  \sqrt{\pi } 
\left(\alpha_3\, \mu \right)^{ \frac{4} {3} }
\right \rbrace }
{99783200 \, \Gamma \left(-\frac{4}{3}\right) 
\Gamma  \left(\frac{5}{6}\right)
\,\mu ^{ \frac{2} {3} } \, \omega^2
}
 \nn & \hspace{ 9 cm } 
 -\frac{1188\, \alpha_3^2\, A_0^6} {775 \, \omega^2 }  
 -\frac{\alpha_3^4\, A_0^{12}} {135 \, \mu ^2\, \omega^2 }
\Bigg ], \nn
& \sigma_{xx}^{(2)} =
\frac{26244 \,e^4\, \tau  \, v_z\, \Gamma \left(\frac{2}{3}\right) 
\Gamma \left(\frac{29}{6}\right) \left(\frac{\alpha_3}
{\mu}\right)^{ \frac{2} {3} } B_x^2 } {150535 \, \pi^{ \frac{5} {2} }}
\, \Bigg[  1 +
\frac{150535\, \sqrt{\pi } \,\alpha_3^{ \frac{10} {3} } \,A_0^{10}}
{3888 \,\mu ^{ \frac{4} {3} } \,\omega ^2 \,
\Gamma \left(\frac{2}{3}\right)\, \Gamma
   \left(\frac{29}{6}\right)}
   +\,\frac{223652 \,\sqrt{\pi }
   \, \alpha_3^{ \frac{4} {3} } \,A_0^4 \,\mu ^{ \frac{2} {3} }}
   {81 \,\omega ^2\, \Gamma\left(\frac{2}{3}\right) 
   \,\Gamma \left(\frac{29}{6}\right)}
   + \frac{4 \,\alpha_3^4 \, A_0^{12}}
   {351 \,\mu ^2 \,\omega ^2}
 \nn &  \hspace{ 7 cm }
  +\frac{8326\,
   \sqrt{3} \,\pi \, \alpha_3^{ \frac{7} {3} } \,A_0^8 \; \Gamma \left(\frac{25}{6}\right) 
 (\alpha_3 \, \mu )^{\frac{1} {3} }}
   {247\, \mu \, \omega ^2\, \Gamma
   \left(\frac{2}{3}\right)^2 
   \, \Gamma \left(\frac{29}{6}\right)}
   +\frac{39690 \,\alpha_3^2\, A_0^6}  {247 \,\omega ^2}
\Bigg ], \nn
& \sigma_{xx}^{(3)} = \frac{ -\, 405\, e^4\, \tau  \, v_z
\, \Gamma \left(\frac{2}{3}\right) \,\Gamma \left(\frac{5}{6}\right) 
\left(\frac{\alpha_3} {\mu}\right)^{\frac{2}{3}} B_x^2 } 
{182 \, \pi ^{\frac{5}{2}}}
 \Bigg [1 +
\frac{182 \,\sqrt{\pi }\, \alpha_3^{\frac{10}{3}} \,A_0^{10}}
{225\, \mu ^{\frac{4}{3}}\, \omega ^2 \,\Gamma \left(\frac{2}{3}\right) 
\,\Gamma
   \left(\frac{5}{6}\right)} +
   \frac{1716 \, \sqrt{\pi } \,\alpha_3^{\frac{4}{3}} \,A_0^4 \,\mu ^{\frac{2}{3}}}
   {25 \,\omega ^2\, \Gamma\left(\frac{2}{3}\right) 
   \, \Gamma \left(\frac{5}{6}\right)}
   -\frac{5 \, \alpha_3^4 \, A_0^{12}}   {513 \,\mu ^2 \, \omega ^2}
\nn & \hspace{ 7 cm}
   + \frac{209573\,\alpha_3^{\frac{7}{3}} \, A_0^8 \, \Gamma \left(\frac{1}{3}\right) 
   \,\Gamma \left(\frac{7}{6}\right) 
   \left( \alpha_3 \,\mu \right)^{\frac{1}{3}} }
   {21505 \,\mu \,\omega ^2 \, \Gamma \left(\frac{2}{3}\right)\,
    \Gamma \left(\frac{5}{6}\right)}
   + \frac{31536 \, \alpha_3^2 \,A_0^6} {475 \,\omega ^2}
\Bigg ] ,
\end{align}
\begin{align}
& \sigma_{yx}^{(1)} =
\frac{4374 \,e^4 \,\tau \, v_z \,\Gamma \left(\frac{11}{6}\right) \,
\Gamma \left(\frac{8}{3}\right) 
\left(\frac{\alpha_3} {\mu}\right)^{\frac{2}{3}} \,B_x  \, B_y} 
{8645 \,\pi ^{\frac{5}{2}}}
\, \Bigg[ 1
- \frac{494 \sqrt{\pi } \alpha_3^{\frac{4}{3}} A_0^4 \,\mu ^{\frac{2} {3} }}
{99 \, \omega ^2 \, \Gamma \left(\frac{11}{6}\right) \,
\Gamma \left(\frac{8}{3}\right)}
-\frac{\alpha_3^4 \,A_0^{12}} {135 \,\mu ^2\, \omega ^2}
+ \frac{2118025 \,\alpha_3^2 \,A_0^8\, 
\Gamma\left(\frac{1}{3}\right) \Gamma \left(\frac{7}{6}\right) \,
 \left( \alpha_3 \,\mu \right)^{\frac{1}{3}} }
 {13470732 \, \mu \, \omega ^2 
 \, \Gamma\left(\frac{11}{6}\right) \, \Gamma \left(\frac{8}{3}\right)}
\nn & \hspace{ 7.5 cm}
 -\frac{1188 \, \alpha_3^2 \, A_0^6}  {775 \,\omega ^2}
\Bigg ],\nn & 
\sigma_{yx}^{(2)} = \frac{B_y} {B_x} \, \sigma_{xx}^{(2)} \,,\quad
\sigma_{yx}^{(3)} = \sigma_{yx}^{(4)} = \frac{B_y} {B_x} \,\frac{ \sigma_{xx}^{(2)}} {2}\,.
\end{align}

Adding up all the parts, we obtain
\begin{align}
&   \sigma_{xx}-  \sigma_{xx} \big \vert_{B=0} \nn
& = 
 e^4 \, \tau\, v_z \, B^2 \left(\frac{\alpha_3}{\mu }\right)^{\frac{2} {3} } 
\Big[ 
0.070099 + 0.049632 \cos (2\theta ) 
+ 
\frac{16.521 \,E_0^4 \,\alpha_3 \,(\alpha_3 \,\mu ^2)^{\frac{1}{3}} 
\left\lbrace \cos (2 \theta)+0.99226 \right \rbrace}         {\omega ^6} 
+
\frac{E_0^6 \,\alpha_3^2 \left\lbrace 13.823 \cos (2 \theta )+13.791 \right \rbrace}
   {\omega ^8}
 \nn & \hspace{ 3.5 cm}  +
\frac{ E_0^8\, \alpha_3^{\frac{8} {3} } \left\lbrace 3.4775 \cos(2 \theta ) + 3.4832 \right \rbrace }
{\mu ^{\frac{2} {3} }\, \omega ^{10}}
   +
\frac{0.50154 \, E_0^{10} \,\alpha_3^{\frac{10} {3} } \cos ^2 \theta }
{\mu ^{\frac{4} {3} } \, \omega ^{12}} 
 + 
\frac{ E_0^{12} \,\alpha_3^4 \left\lbrace 0.0022362 \cos (2 \theta) + 0.0020846 \right \rbrace }
{\mu ^2 \,\omega ^{14}} 
\Big ]
\end{align}
and
\begin{align}
   \sigma_{yx}  &=
e^4 \,\tau \, v_z \,B^2 \sin (2 \theta ) \left(\frac{\alpha _3}{\mu }\right)^{ \frac{2} {3}}
\Big[ 0.049632 
+\frac{16.521 \, E_0^4 \,\alpha _3^{ \frac{4} {3}} \, \mu ^{ \frac{2} {3}}}{\omega ^6}
+ \frac{13.823 \,E_0^6 \, \alpha _3^2} {\omega ^8}
+ \frac{3.4775 \,E_0^8 \,\alpha _3^{ \frac{8} {3}}} 
{\mu ^{ \frac{2} {3}} \,\omega^{10}}
+ \frac{0.25077 \,E_0^{10} \, \alpha _3^{ \frac{10} {3}}}  {\mu ^{ \frac{4} {3}} \,\omega ^{12}}
\nn & \hspace{ 4.75 cm}
+ \frac{0.0022362 \, E_0^{12}\, \alpha _3^4}
   {\mu ^2 \omega^{14}} 
\Big ].
\end{align}

\begin{figure}[t]
	\centering
{\includegraphics[width= 0.75 \textwidth]{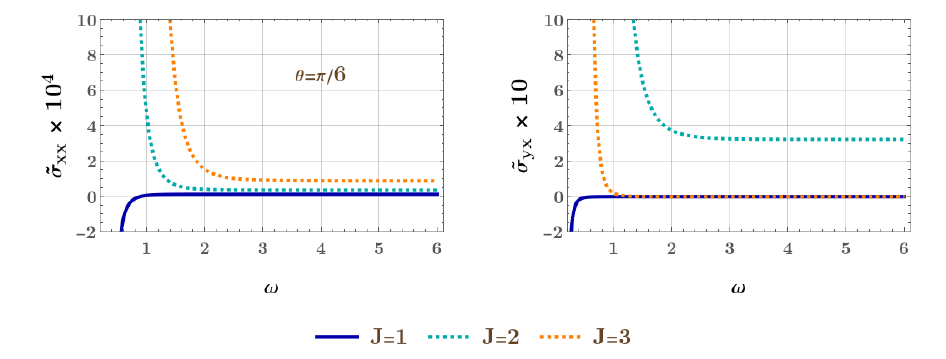}}
\caption{\label{figsig}
Left panel: Plots for $\tilde \sigma_{xx}$ (in the units of eV$^{-2}$) as a function of $\omega$, where $\tilde \sigma_{xx}
=\left ( \sigma_{xx} -\sigma_{xx} \vert_{B=0} \right ) / (e^2 \,\tau \, v_z \,B^2)$, setting $\theta = \pi/6 $. Right panel: Plots for $\tilde \sigma_{yx}$ (in the units of eV$^{-2}$) as a function of $\omega$, where $\tilde \sigma_{yx}
= \sigma_{yx} / (e^2 \,\tau \, v_z \,B^2 \sin \theta \cos \theta)$. The values of $ \mu $, $\alpha_J$'s, and $E_0$ are taken from Table~\ref{table_quantites}.}
\end{figure}

\subsection{Magneto-thermoelectric conductivity: LTEC and TTEC}
\label{secthermo}

Using Eq.~\eqref{eqsigma}, the LTEC and the TTEC are given by 
\begin{align}
  \alpha_{xx} & 	\approx e \, \tau
\int\frac{d^{3} \boldsymbol k} {(2 \, \pi)^{3}} \,\mathcal{D}
\left \lbrace  {v_x}+{e\,B_x}
\left (\boldsymbol{v}\cdot {\boldsymbol{\Omega}}_F^{(0)}  \right ) \right \rbrace^2
\frac{\left (\mu-\epsilon_{\boldsymbol k} \right )  } {T}
\left(-\frac{\partial f_0}
{\partial\epsilon_{\boldsymbol{k}}}\right),	
\text{ and}		
\nn   \alpha_{yx}& \approx e \, \tau
	\int\frac{d^{3} \boldsymbol k} {(2 \, \pi)^{3}} \,\mathcal{D}
  \left(-\frac{\partial f_0} {\partial\epsilon_{\boldsymbol{k}}}\right) 
\left\lbrace v_y + e\,B_y
\left (\boldsymbol{v }\cdot {\boldsymbol{\Omega}}_F^{(0)}   \right )
\right \rbrace
\left\lbrace v_x + e\,B_x
\left (\boldsymbol{v }\cdot {\boldsymbol{\Omega}}_F^{(0)}  \right ) \right \rbrace
\frac{\left (\mu-\epsilon_{\boldsymbol k} \right )  } {T}
\left(-\frac{\partial f_0}
{\partial\epsilon_{\boldsymbol{k}}}\right),
\label{eq_ttec_ltec}
\end{align}
respectively. 
On the other hand, assuming a small-temperature limit of $T\ll \mu $, we can apply the Mott relations, which tell us that
\begin{align}
\partial_{ \mu }   \sigma_{ij} 
= - \,\frac  {3\, e }  {\pi^2 \, T} \, \alpha_{ij} + \order{T^2}\,.
\label{eqmott}
\end{align}
Therefore, instead of explicitly evaluating the integrals leading to extremely lengthy and cumbersome expressions, one can invoke the Mott relations to infer the characteristics of the LTEC and the TTEC. We would like to point out that the explicit expressions in the static limit can be found in Refs.~\cite{ips_rahul_ph_strain, ips-ruiz}, where they are shown to satisfy the Mott relations.

\subsection{Magnetothermal coefficients}
\label{secthermal}

Using Eq.~\eqref{eqsigma}, the in-plane components of the magnetothermal conductivity are given by 
\begin{align}
  \ell_{xx} & 	\approx   \tau
\int\frac{d^{3} \boldsymbol k} {(2 \, \pi)^{3}} \,\mathcal{D}
\left \lbrace  {v_x}+{e\,B_x}
\left (\boldsymbol{v}\cdot {\boldsymbol{\Omega}}_F^{(0)}  \right ) \right \rbrace^2
\frac{\left (\epsilon_{\boldsymbol k} -\mu \right )^2  } {T}
\left( -  \frac{\partial f_0}
{\partial\epsilon_{\boldsymbol{k}}}\right)	
\text{ and}		
\nn  \alpha_{yx}& \approx  \tau
	\int\frac{d^{3} \boldsymbol k} {(2 \, \pi)^{3}} \,\mathcal{D}
  \left(\frac{\partial f_0} {\partial\epsilon_{\boldsymbol{k}}}\right) 
\left\lbrace v_y + e\,B_y
\left (\boldsymbol{v }\cdot {\boldsymbol{\Omega}}_F^{(0)}   \right )
\right \rbrace
\left\lbrace v_x + e\,B_x
\left (\boldsymbol{v }\cdot {\boldsymbol{\Omega}}_F^{(0)}  \right ) \right \rbrace
\frac{\left ( \epsilon_{\boldsymbol k} -\mu \right )^2  } {T}
\left(-\frac{\partial f_0}
{\partial\epsilon_{\boldsymbol{k}}}\right),
\label{eq_ttec_ltec}
\end{align}
respectively. On the other hand, assuming a small-temperature limit of $T\ll \mu $, we can apply the relation embodied by
\begin{align}
 \sigma_{ij} =\frac{ 3 \,e^2 } {\pi^2\, T}   \,   \ell_{ij} + \order{T^2} ,
\label{eqwf}
\end{align}
which holds by the virtue of the Wiedemann-Franz law. Again, this law is applicable in the limit $T \rightarrow 0 $ \cite{mermin}. Since $\ell $ is related to thermal conductivity ($\kappa$) \cite{mermin, ips-kush-review, ips-hermann-review, ips-hermann-thermo}, the characteristics of the in-plane components of $\kappa$ can also be inferred from our explicit derivations of the in-plane components of $\sigma$.

\subsection{Discussions of the nature of the resulting conductivity}
\label{sec_dis_con}
 
\begin{figure}[t]
	\centering
{\includegraphics[width= 0.5 \textwidth]{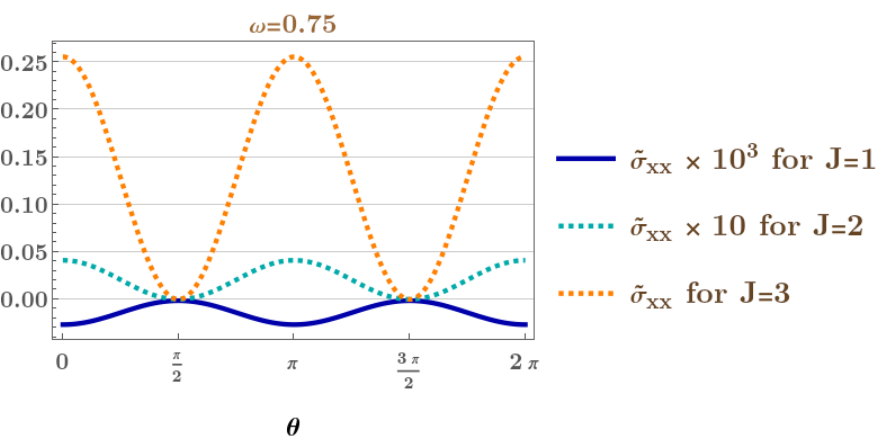}}
\caption{\label{figsig2}
Plots for $\tilde \sigma_{xx}$ (in the units of eV$^{-2}$) as a function of $ \theta$, where $\tilde \sigma_{xx}
=\left ( \sigma_{xx} -\sigma_{xx} \vert_{B=0} \right ) / (e^2 \,\tau \, v_z \,B^2)$, setting $ \omega = 0.75$ eV. The values of $ \mu $, $\alpha_J$'s, and $E_0$ are taken from Table~\ref{table_quantites}.}
\end{figure}

In this subsection, we will discuss the results obtained for the transport coefficients. 
For each value of $J$, we note that there appears a Drude term in $\sigma_{xx}^{(1)}$, which is proportional to $\mu^2$ in the static limit. It refers to the part independent of $\boldsymbol B$. However, no nonzero Drude term exists for the in-plane transverse components. With the choice of parameter values listed in Table~\ref{table_quantites}, with which the validity of the semiclassical Boltzmann transport theory is justified, both the LMC and the PHC gravitate towards the static case values (which can be found in Refs.~\cite{Nag_2020, ips_rahul_ph_strain}) quite fast, as $\omega $ is cranked up. This is regardless of the topological charge $J$, which can be traced to the fact that the leading-order-in-$\omega^{-1}$ correction falls off as $ E_0^4/\omega^ 6$. Consequently, the dependencies on the parameters (viz. $\mu$ and $B$) in the static case continue to dictate the dominant behaviour, even after including the time-periodic drive. For the $\boldsymbol B$-dependent terms, they include the effects of the BC and show a strong dependence on $J$, because the zeroth-order term (pertaining to the static case~\cite{Nag_2020, ips_rahul_ph_strain}) scales as $\mu^{-2/J}$. The overall behaviour as a function of $\omega$ is illustrated in Fig.~\ref{figsig}. 

Apart from the nonzero Drude part of $\sigma_{xx}$, everything else has a quadratic-in-$B$ dependence, which is expected on account of the Onsager-Casimir reciprocity relation, $\sigma_{ij} (\mathbf{B})= \sigma_{ji} (-\mathbf{B})$ \cite{onsager1, onsager2, onsager3}. Stated more explicitly, when the system is subjected to homogeneous external fields, in the absence of any other scale in the problem, the Onsager-Casimir reciprocity relation forbids any term from being linear-in-$B$. This holds unless the change of sign of $\mathbf B $ is compensated by a change of sign in another parameter in the system, for example, a tilt in the spectrum \cite{ips-rahul-jpcm} or the emergence of an axial pseudomagnetic field \cite{ips_rahul_ph_strain}.
Hence, our conclusion here is that all the $B$-dependent parts of the conductivity tensor arise as $B_x^2$ or $B_x\, B_y$, elucidating the fact that they are all sinusoidal functions of $(2\theta)$. Consequently, the resulting curves have a $\pi$-periodic dependence on $\theta$. Fig.~\ref{figsig2} illustrates the variations of the longitudinal components with respect to $\theta$, which is provided mainly to compare the magnitude of the response for the three values of $J$, plugging in some typical parameter values. We find that the magnitude increases with increasing $J$. In light of the Mott relations and the Wiedemann-Franz law [cf. Eqs.~\eqref{eqmott} and \eqref{eqwf}], it is obvious that the behaviour of the components of $\alpha $ and $\ell$ should follow suit. 

 \section{Summary and outlook}
\label{secsum}

In this paper, we have evaluated the components of the in-plane conductivity for WSMs/mWSMs in planar-Hall set-ups, when the system is perturbed by a high-frequency time-periodic drive. From there, we have inferred the behaviour of the thermoelectric coefficients, applicable for the planar-thermal-Hall configurations.
We have used the low-energy effective Hamiltonian for a single node and, using a combination of the Floquet formalism and van Vleck's perturbation theory, we have obtained the leading-order corrections in the high-frequency limit. This serves as a complementary signature for these semimetallic systems, in addition to studies of other transport properties (see, for example, Refs.~\cite{Nandy_2017, ips-tunnel-qbcp, ips-tunnel-qbcp-delta, Nag_2020, ips-trans, Nag_floquet_2020, ips_aritra, sajid_cd, sajid_magnus, ips-jns}) of WSMs/mWSMs. In particular, the topological charge, equalling $J$, etches a unique signature on the linear-response tensors arising in the planar-Hall set-ups, through the BC terms appearing in the integrands.
Additionally, the periodic drive provides an extra control knob, in terms of the frequency-dependence of the incident circularly-polarized electromagnetic fields, thus making the overall transport coefficients depend on the parameter $\omega$.

We note that the high-frequency (or large $\omega$) regime implies a \text{weak} perturbative effect from the driving part, compared to the original unperturbed Hamiltonian [cf. Eq. \eqref{eq_multi1}]. This is clear from the fact that $\omega $ appears in the denominator (and not in the numerator). Furthermore, the strength of the drive is determined by the amplitide $A_0= {e \,E_0} / {\omega}$ [see the lines before Eq. \eqref{h1_time}]. Being a perturbative part, the periodic drive's effect in modifying topological properties is also subleading, which is captured in the modified BC shown in Eq. \eqref{eq_bcl}. All our answers are obtained as corrections in powers of $A_0$, as a consequence of Eqs. \eqref{v1} and \eqref{h5}. In summary, high frequency and weak driving-strength are synonymous here.

While computing the analytical expressions for the magneto-transport coefficients, we have limited ourselves to the large-frequency limit [captured by Eq.~\eqref{eqh4}] and have shown the corrections by expanding the physical quantities up to order $\lambda^2$. Using the systematic scheme that we have outlined, higher-order corrections can also be incorporated into the expressions, if desired. Our results show that, for large but finite $\omega$, the behaviour of the ($\omega$-dependent) perturbative terms are significantly different with respect to the static case (corresponding to $\omega^{-1} \rightarrow 0 $). As is evident from our results, these pertubative terms are complicated functions of the chemical potential ($\mu$) and the material-dependent parameter $\alpha_J$. It is to be noted that these parameters enter into the static parts via $(\alpha_J / \mu)^{2/J}$ \cite{Nag_2020, ips_rahul_ph_strain}. The difference from the static values amplifies especially for small values of $\mu$, as some of the terms in the analytical expressions contain inverse powers of $\mu$. Last but not the least, the transport characteristics reported here comprise an important probe for the Berry curvature (BC), which constitutes unconventional properties involving topological phases beyond the conventional symmetry-breaking paradigm of Ginzburg-Landau. The BC here varies with $J$, thereby reflecting the inherent topological properties of the nodal points --- namely, whether it is a Weyl, double-Weyl, or triple-Weyl node. Thus, our results serve as a guide to what signatures to look for in experiments, while measuring transport-properties like conductivity.

Let us elaborate on a viable platform where the results captured by our theoretical expressions can be explicitly tested experimentally. Using a material such as $\text{HgCr}_{2}\text{Se}_{4}$ as the sample, the Floquet driving can be incorporated by the pump-probe spectroscopy, which is the simplest experimental technique used to study ultrafast electronic dynamics. Therein, an ultrashort laser pulse is split into two portions --- (1) a beam (pump) is used to excite the sample, generating a short-lived non-equilibrium state; and (2) a beam (probe) is used to monitor the pump-induced changes in the optical constants (such as reflectivity and transmission) of the sample.
The delay between the pump- and probe-pulses is adjusted by changing the path-length difference between them, inducing a relative time-delay. By monitoring the probe signal with a detector, after the probe pulse has interacted with the sample as a function of the time-delay, it is possible to obtain information on the decay dynamics of the generated excitation. The modulations of the probe signal constitute the key to deducing the dynamics of the electronic energy levels, as they decay back to the equilibrium states after the perturbation caused by the pump-pulse. The changes in the optical constants, for example, as functions of the time-delay between the arrival of pump- and probe-pulses provide valuable information about the relaxation-processes of electronic states in the sample. 
Indeed, by shining circularly-polarized photons at mid-infrared wavelengths, Floquet-Bloch states have been demonstrated to exist on the surface of a topological insulator \cite{expt-pump-probe}. Such an arrangement thus serves as a protoype for designing set-ups for WSMs/mWSMs, with suitably chosen frequency-ranges of the pump laser, for measuring the response coefficients derived here.

The inclusion of tilt in the spectrum captures generic situations involving WSMs/mWSMs.
In particular, tilting causes linear-in-$B$ terms to appear in the in-plane response coefficients, as found in Refs.~\cite{ips-rahul-jpcm, ips-tilted, ips-shreya, ips_tilted_dirac}.
The linear-response coefficients in the presence of a quantizing magnetic field have been computed for (1) tilted WSMs/mWSMs in Ref.~\cite{ips_ll} and (2) 2d semi-Dirac semimetals in Ref.~\cite{ips-kush}. 
It will be worthwhile to compute the effects of a high-frequency periodic drive for all these cases. A challenging avenue is to carry out the conductivity calculations in the presence of interactions (such that when interactions affect the quantized physical observables in the topological phases \cite{kozii,Mandal_2020}, or when non-Fermi liquids emerge \cite{ips-hermann-cond,ips-hermann-thermo,ips-raman}), and/or disorder \cite{rahul-sid,ips-rahul,ips-klaus,ips-biref}.
Application of Coulomb interactions, for example, can induce plasmons, whose salient features can be computed in lines with the treatment in Refs.~\cite{ips-plasmons, ips-jing-plasmons}. In many cases, long-ranged (unscreened) Coulomb interactions, with the chemical potential fine-tuned to cut exactly the nodal point, can induce strongly-correlated non-Fermi liquids \cite{Abrikosov, moon-xu, ips-hermann-cond, ips-hermann-thermo} or marginal Fermi liquids \cite{malcolm-bitan, ips-biref}, which fall outside the paradigm of Landau's Fermi-liquid description. Regarding disorder, weak disorder actually act as the centres of collision which cause scattering and relaxation processes in scenarios like those considered here. However, very strong disorder can have drastic effects which require incorporation of many-body techniques \cite{rahul-sid, ips-rahul, ips-klaus,ips-biref}. What the outcomes of these might be have to be computed on a case-by-case basis.


\appendix 

\section*{Appendix: Detailed steps to evaluate the integrals for LMC and PHC}
\label{appsig}

In this appendix, we will outline the computation of the terms shown in Eq.~\eqref{eq_lmc_break}.
For the ease of carrying out the integrals, we employ a change of variables via the following transformations:
\begin{align}
\label{eqsph} 
k_x =  \left( \frac{ \varepsilon}{\alpha_J } \sin \gamma \right)^{1/J} \cos \phi \,, \quad
k_y =  \left( \frac{ \varepsilon}{\alpha_J } \sin \gamma \right)^{1/J} \sin \phi\,, 
\quad k_{z} = \frac{ \varepsilon}{v_z } \cos \gamma\,,
\end{align}
where $ \varepsilon \in [0, \infty )$, $\phi \in [0, 2 \pi )$, and $\gamma \in [0, \pi ]$. 
The Jacobian for the transformation is
$\mathcal{J} ( \varepsilon , \gamma ) =   \frac{1} {J\, v_z \sin \gamma } 
\left( \frac{ \varepsilon \sin \gamma}
{\alpha_J } \right) ^{\frac{2}{J}}$.

Let us re-express Eq.~\eqref{eff_eng} for the quasi-energies as
\begin{align}
	\epsilon_{\boldsymbol{k}}   & =  
\sqrt{ \varepsilon^2 
+   \lambda \, \frac{ 2\, \varepsilon \cos \gamma\, 
 \chi_J ( \varepsilon, \gamma) } {\omega}
+  \lambda^2 \, \frac{\chi_J^2 ( \varepsilon, \gamma)} {\omega^2}
}\,, \text{ where }
\chi_J ( \varepsilon, \gamma) = 
( \varepsilon \sin{\gamma})^2 
\sum \limits_{p=1}^{J} \frac{(^J C_p \,A_0^p)^2} {p}
\left( {\frac {\alpha_J}  { \varepsilon \sin{\gamma}}}\right)^{\frac{2\,p}{J}} .
\label{eq_ene_def}
\end{align}
Since we want to limit ourselves to $\lambda^2 $, we can approximate the quasi-energy as 
\begin{align}
\label{eq_ene_def}
	\epsilon_{\boldsymbol{k}}  =  \varepsilon + \tilde{\chi}_J (\varepsilon, \gamma)
	+ \mathcal{O} \left( \lambda^3 \right),
	\text{ where }
\tilde{\chi}_J (\varepsilon, \gamma) = 
 \lambda \, \frac{ \chi_J (\varepsilon, \gamma) \cos \gamma} {\omega} 
+ \frac{\lambda^2} {\omega^2} \, \frac{\chi_J^2 (\varepsilon, \gamma) \sin^2 \gamma} {2\,\varepsilon} .	
\end{align}

Aided by the expressions in Eqs.~\eqref{eq_vals} and \eqref{eq_ene_def}, we first expand the integrands in Eqs.~\eqref{s0} and \eqref{syx} in small $B$, in order to take care of the $\boldsymbol B$-dependent terms in the phase-space factor $\mathcal D$. Retaining terms upto $\mathcal{O}(B^2)$, we perform the $\phi$-integrals as the second step. After that step, each integral reduces to a generic form as shown below:
\begin{align}
	I_{\gamma, \varepsilon} = \int d\gamma \,  d \varepsilon  \; F ( \varepsilon, \mu, \gamma ) \; 
	\delta \Big(  \varepsilon + \tilde \chi_J(\varepsilon, \gamma)  -  \mu \Big) \,,
\end{align}
where $ F ( \varepsilon, \mu, \gamma )$ is a function of $ \varepsilon$, $\mu$, and $\gamma $.
We now implement the expansion
\begin{align}
\delta \big(  \varepsilon + \chi_J(\varepsilon, \gamma) \cos \gamma -  \mu \big) 
=
\delta \big(  \varepsilon -  \mu \big) + \tilde  \chi_J(\varepsilon, \gamma) 
 \; \partial_{\varepsilon} [\delta \big(  \varepsilon -  \mu \big)]
 +  \mathcal{O} \left(\frac{1}{\omega^2} \right).
\end{align}
Thus, retaining the leading-order corrections in $ \omega^{-1} $, we arrive at
\begin{align}
	I_{\gamma, \varepsilon} =  
	\int_0^{\pi} d\gamma \Big[\, F ( \varepsilon, \mu, \gamma ) 
- 
\partial_\varepsilon \left  \lbrace 
F ( \varepsilon, \mu, \gamma ) \, \tilde \chi_J (\varepsilon,\gamma) 
 \right \rbrace \Big ]_{ \varepsilon = \mu}  .
\end{align}
We now use this form to evaluate the final integrals and to get the final expressions shown in the main text.


\bibliography{ref}

\end{document}